\documentclass[aps,usletters,prl,twocolumn,showpacs,superscriptaddress,floatfix]{revtex4-1}
\usepackage[dvips]{graphicx,color,colortbl}
\usepackage[english]{babel}
\usepackage{enumerate,amsmath,amssymb,amsbsy,amscd,bm}

\newcommand{\ket}[1]{|{#1} \rangle}  
\newcommand{\bra}[1]{\langle {#1}|}  
\newcommand{\ketbra}[2]{|{#1}\rangle\langle{#2}|}  

\begin{document}
\preprint{}
\title[]{Non-Adiabatic Quantum Dynamics of Grover's Adiabatic Search Algorithm}
\author{Sangchul Oh}
\email{email: oh.sangchul@gmail.com}
\affiliation{Qatar Environment and Energy Research Institute, Qatar Foundation, Doha, Qatar}
\author{Sabre Kais}
\email{email: kais@purdue.edu}
\affiliation{Qatar Environment and Energy Research Institute, Qatar Foundation, Doha, Qatar}
\affiliation{Department of Chemistry, Department of Physics and Birck Nanotechnology Center,
Purdue University, West Lafayette, IN 47907 USA}
\date{\today}

\begin{abstract}
We study quantum dynamics of Grover's adiabatic search algorithm with the equivalent two-level 
system. Its adiabatic and non-adiabatic evolutions are visualized as trajectories of Bloch 
vectors on a Bloch sphere. We find the change in the non-adiabatic transition probability 
from exponential decay for short running time to inverse-square decay for long running time. 
The size dependence of the critical running time is expressed in terms of Lambert $W$ function. 
The transitionless driving Hamiltonian is obtained to make a quantum state follow the adiabatic 
path. We demonstrate that a constant Hamiltonian, approximate to the exact time-dependent driving 
Hamiltonian, can alter the non-adiabatic transition probability from the inverse square decay to 
the inverse fourth power decay with running time. This may open up a new way of reducing errors in 
adiabatic quantum computation.
\end{abstract}
\maketitle

 
\paragraph{Introduction--}Grover's quantum search algorithm~\cite{Grover97} is known to find a 
marked one out of $N$ entries with the $O(\sqrt{N})$ queries on a quantum computer, otherwise 
the $O(N)$ queries are needed on a classical computer. While it was initially 
designed to be implemented on a quantum circuit model, its adiabatic quantum computation 
version~\cite{Farhi}, called Grover's adiabatic search algorithm, was also proposed and the equivalence 
between them was proved~\cite{Vandam,Aharonov,Roland03}.

Much attention has been paid to solving an instantaneous eigenvalue problem of a time-dependent 
Hamiltonian of an adiabatic quantum algorithm because the minimum gap of a system determines 
the validity of an adiabatic quantum evolution and thus its computational 
complexity~\cite{Messiah,Farhi}. The non-adiabatic transition to other states, i.e., the deviation 
from the adiabatic evolution, is the main concern in adiabatic quantum computation. To know in 
detail how the non-adiabatic transition decreases asymptotically with running time, the minimum 
gap of the instantaneous eigenvalues is not enough, so a time-dependent Schr\"odinger equation
has to be solved. 

In this paper, we study quantum dynamics of Grover's adiabatic search algorithm with an equivalent 
two-level system to calculate its non-adiabatic transition probability. The adiabatic and
non-adiabatic evolutions of a quantum state are represented by trajectories on a Bloch sphere.
We show that the non-adiabatic transition probability changes from exponential decay
for short running time to inverse square decay for long running time. The dependence of 
the critical running time on the problem size is written in terms of Lambert $W$ function. 
Finally, We show that a constant driving Hamiltonian could reduce significantly 
the non-adiabatic transition probability, which may speed up adiabatic quantum computation.

 
\paragraph{Hamiltonian of adiabatic search algorithm--} 
Let us start with introducing the time-dependent Hamiltonian for Grover's adiabatic search 
algorithm~\cite{Roland03,Schaller}. The adiabatic quantum computation is based on the adiabatic 
theorem which states that if a time-dependent Hamiltonian changes slowly enough, then an eigenstate 
of an initial Hamiltonian, an input state, evolves to an eigenstate of a final Hamiltonian, an output 
state~\cite{Farhi,Messiah}. Grover's search algorithm takes the input state as a superposition of 
all possible states $\ket{\varphi_{\rm in}} = \frac{1}{\sqrt{N}}\sum_{i=0}^{N-1}\ket{i}$ 
with $N$ entries.  It is the ground state of the initial Hamiltonian 
$H_0 = \mathbf{I} -\ketbra{\varphi_{\rm in}}{\varphi_{\rm in}} 
    = \mathbf{I} - \frac{1}{N}\sum_{i,j}\ketbra{i}{j}$
where $\mathbf{I}$ is an $N\times N$ identity matrix. Note that 
$\sum_{i,j}\ketbra{i}{j}$ is a matrix with all entries 1 whose eigenvalues are 0 ($N-1$ multiples)
and $N$~\cite{Horn}. The output or target state $\ket{w}$ to find is the ground 
state of the final (or problem) Hamiltonian $H_p = \mathbf{I} -\ketbra{w}{w}$.
The slow change from the initial to final Hamiltonians can be done as
$H(t) = f(s)H_0 + g(s)H_p$
where $s\equiv t/T$ is the dimensionless (or macroscopic) time~\cite{Messiah,Betz}, 
$T$ is the running time acting as an adiabatic parameter, and a turn-off function $f(s)$ 
and turn-on function $g(s)$ satisfy $f(0) = g(1) =1$ and $f(1) = g(0) = 0$. 
The simplest choice of $f$ and $g$ is to interpolate $H_0$ and $H_p$ linearly, i.e., 
$f(s) = 1-s$ and $g(s) = s$.

\paragraph{Instantaneous eigenvalues and eigenstates--} 
Grover's search algorithm is understood as a rotation from the input state $\ket{\varphi_{\rm in}}$ 
to the target state $\ket{w}$. This implies it is essentially a two-dimensional problem formed by
two linearly-independent vectors $\ket{\varphi_{\rm in}}$ and $\ket{w}$. While in quantum circuit 
model the full rotation is done by $O(\sqrt{N})$ successive finite rotations, it is done by 
a continuous rotation in adiabatic quantum computation. The two vectors $\ket{w}$ and 
$\ket{\varphi_{\rm in}}$ are linearly independent but not orthogonal. An orthonormal basis is easily 
constructed from the matrix representation of 
$\mathbf{I} -H(s) = f(s)\ketbra{\varphi_{\rm in}}{\varphi_{\rm in}} + g(s)\ketbra{w}{w}$ whose 
only the $w$-th diagonal element is different. Thus, the time-dependent Hamiltonian for 
Grover's adiabatic search algorithm is represented with the orthonormal basis 
$\{\ket{w},\ket{w_\perp}\}$ as
\begin{align}
H(s)= \mathbf{I} - \frac{f}{N} 
 \left[ \begin{array}{cc}
 1 +N\frac{g}{f} & \sqrt{N-1}\\[10pt]
 \sqrt{N-1}           & N-1
 \end{array}\right]\,,
\label{Hamil_B}
\end{align}
where $\ket{w_\perp} = \frac{1}{\sqrt{N-1}}\sum_{i\ne w}^{N} \ket{i} \,.$
Hamiltonian~(\ref{Hamil_B}) is written in convenient form as 
\begin{align}
H(s) = \frac{(f+g)}{2}\,\mathbf{I}
    -\frac{1}{2N}\left[\begin{array}{rr}
       Z(s)  & X(s) \\[10pt]
       X(s)  & -Z(s)
      \end{array}\right]\,,
\label{Hamil_C}
\end{align}
where $Z(s)\equiv 2f +N(g -f)$  and $X(s) \equiv 2f\sqrt{N-1}$. Since the first term in 
Eq.~(\ref{Hamil_C}) is not relevant to dynamics, it will be dropped.
The Hamiltonian is written as
\begin{align}
H(s) =
    -\frac{\hbar\omega(s)}{2}\left[\begin{array}{rr}
       \cos\theta(s)  & \sin\theta(s) \\[10pt]
       \sin\theta(s)  &-\cos\theta(s)
      \end{array}\right]\,,
\label{Hamil_D}
\end{align}
where the gap between the ground and excited states is given by 
$\hbar\omega(s) \equiv\frac{1}{N} \sqrt{Z^2 + X^2} =  \sqrt{ (f-g)^2 + \frac{4}{N}fg}\,$.
Here mixing angle $\theta$ is defined by $\tan\theta(s) \equiv X(s)/Z(s)$. 
While a different choice of $f$ and $g$ gives rise to a different energy gap, 
hereafter we consider only a linear interpolation case.
Hereafter we set $\hbar=1$.

\begin{figure}[htbp]
\includegraphics[scale=1.0, angle=0]{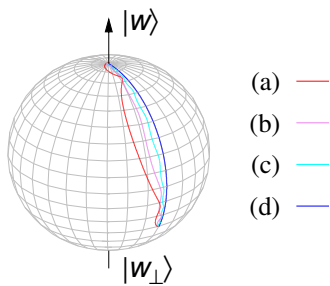}
\caption{(color online). Trajectories of a Bloch vector ${\bf r}(t)$ on a Bloch sphere 
for various running times (a) $T=10$, (b) $T=100$, (c) $T=300$. (d) The blue longitudinal 
line represents the adiabatic path. Here $N=4$ is set. If $N$ is large, 
an initial Bloch vector becomes closer to $\ket{w_\perp}$.}
\label{Fig1}
\end{figure}

As in a textbook of quantum mechanics, the instantaneous eigenstates of 
$H(s)\ket{e_\pm(s)} = e_\pm(s) \ket{e_\pm(s)}$ read
\begin{align}
\ket{e_{-}(s)} 
= \left[\begin{array}{cc}
 \cos\tfrac{\theta}{2}\\[10pt]
 \sin\tfrac{\theta}{2}
 \end{array}\right]\,,\quad
\ket{e_{+}(s)} 
= \left[ \begin{array}{rr}
  -\sin\tfrac{\theta}{2}\\[10pt]
   \cos\tfrac{\theta}{2}
  \end{array}\right]\,.
\end{align}
As represented by a Bloch vector in Fig.~\ref{Fig1}, the input state 
$\ket{\varphi_{\rm in}}=\ket{e_{-}(0)}$ is a vector with azimuthal angle 
$\tan\theta=(2-N)/2\sqrt{N-1}$. The target state $\ket{w} = \ket{e_-(1)}$ points to the north 
pole. Thus, like the Landau-Zener-Majorana-St\"uckelberg 
problem~\cite{Landau,Zener,Majorana,Stuckelberg}, Grover's adiabatic search algorithm is just 
a rotation of a single qubit driven by time-dependent Hamiltonian~(\ref{Hamil_D}). 

\paragraph{Quantum dynamics of adiabatic search algorithm--} 
To understand non-adiabatic effects, we solve numerically a time-dependent Schr\"odinger equation 
\begin{align}
i\hbar\frac{d}{dt}\ket{\psi(t)} = H_T(t) \ket{\psi(t)}\,,
\end{align}
where a time-dependent Hamiltonian $H_T(t)$ is given by Eq.~(\ref{Hamil_D}). 
As illustrated in Fig.~\ref{Fig1}, a quantum state $\ket{\psi(t)} = \alpha(t)\ket{w} + 
\beta(t) \ket{w_\perp}$ is visualized by  a Bloch vector 
${\bf r}(t) \equiv \bra{\psi(t)}\bm{\sigma}\ket{\psi(t)}$ with Pauli matrices 
$\sigma_k$ for $k=x,y,z$. In the adiabatic limit of $T\gg \sqrt{N}$, an evolved quantum state 
remains in the instantaneous ground state, that is, $\ket{\psi(t)} \simeq \ket{e_{-}(t)}$ up to 
the dynamical and geometric phase factors. So, the Bloch vector 
${\bf r}_{\rm ad}(s) = \bra{e_{-}(s)}\bm{\sigma}\ket{e_{-}(s)}$ travels to the north pole 
along the longitude line on a Bloch sphere.

\begin{figure}[htbp]
\includegraphics[scale= 1.0,angle=0]{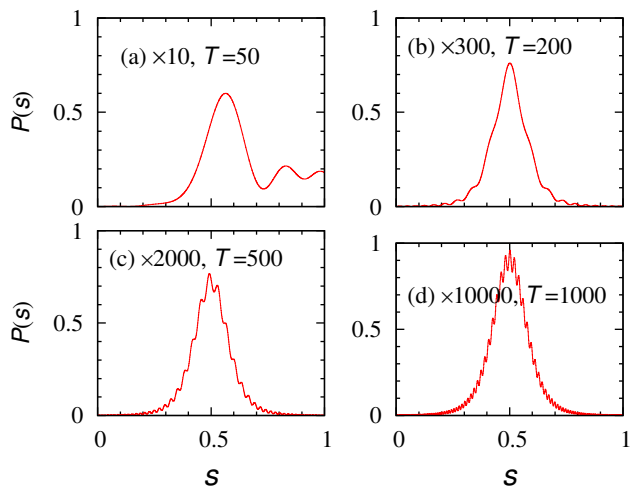}
\caption{(color online). Transition probability $P(s)$ of being in an instantaneous eigenstate 
$\ket{e_{+}(s)}$ as a function of $s$ for (a) $T = 20$, (b) $T=200$, (c) $T=500$, (d) $T=1000$. 
Here the system size $N=10$ is taken. $P(s)$ is magnified by $10,300,2000,10000$ times, 
respectively.}
\label{Fig_dev}
\end{figure}

The adiabatic path is a good approximation to the exact evolution if running time $T$ is large 
enough, that is, the Hamiltonian changes slowly enough. For finite running time $T$, however, 
a real path deviates from the adiabatic path as illustrated in Fig.~\ref{Fig1}. 
To see this in detail, we examine how a quantum state $\ket{\psi(t)}$ is deviated
from the instantaneous eigenstate $\ket{e_{-}(t)}$ as adiabatic parameter $T$ is varied. 
The evolved state $\ket{\psi(t)}$ is written
in terms of instantaneous eigenstates as $\ket{\psi(s)} = a(s)\ket{e_{-}(s)} + b(s) \ket{e_{+}(s)}\,$.
Fig.~\ref{Fig_dev} plots the transition probability $P(s) = 1- |a(s)|^2$ of being in 
an instantaneous ground state $\ket{e_{+}(s)}$ for various running time $T$. For short running 
time $T$, as shown in Fig.~\ref{Fig_dev}~{(a)}, the maximum of $P(s)$ does not coincide with 
the location of the minimum energy gap.  As depicted in Figs.~\ref{Fig_dev}~{(b), (c), and (d)}, 
$P(s)$ becomes smaller and more symmetric and reaches at its peak at $s=1/2$ 
as running time $T$ is increased.

\begin{figure}[ht]
\includegraphics[scale = 1.0,angle=0]{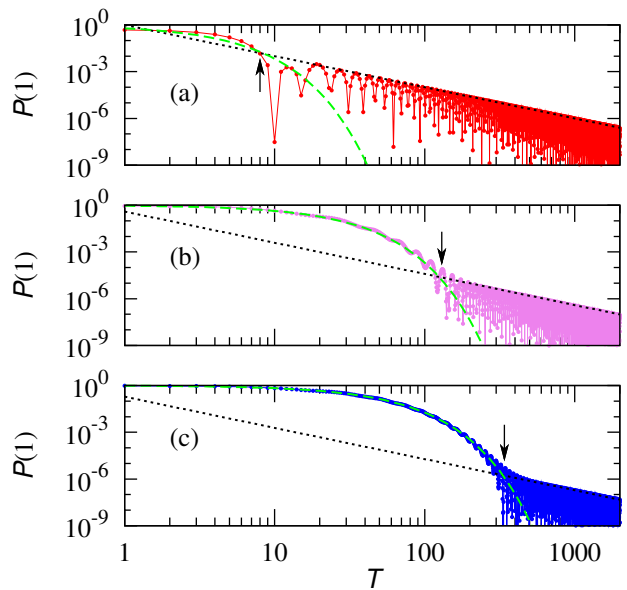}
\caption{(color online). Log-log plots of the non-adiabatic transition probability $P$ as
function of running time $T$ for (a) $N=$ 2, (b) $N=10$, and (c) $N=20$.
The green dashed line is $\exp(-AT)$ and the black dotted line is $B/T^2$. 
The arrows indicate the critical running time $T_c$.}
\label{Fig3}
\end{figure}

\begin{figure}[ht]
\includegraphics[scale = 1.0,angle=0]{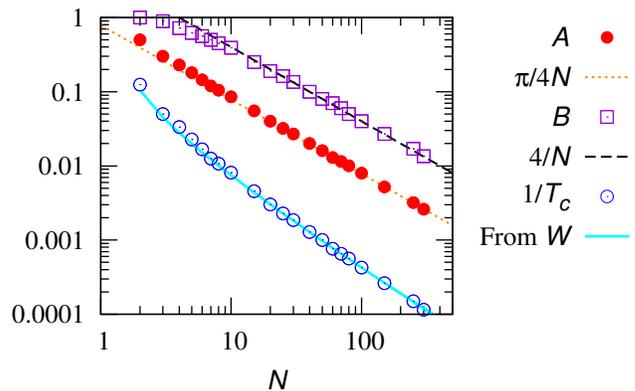}
\caption{(color online). Coefficients $A$ and $B$ in Eq.~(\ref{transition_prob}), and $1/T_c$ 
as a function of the system size $N$. The cyan solid line is a plot with Eq.~(\ref{Tc_Lambert}).}
\label{Fig4}
\end{figure}

\paragraph{Transition of non-adiabatic transition--} The non-adiabatic transition probability 
$P(1)$ at $s=1$ indicates the error of adiabatic quantum computation. The asymptotic form of $P(1)$ 
for the Landau-Zener-Majorana-St\"uckelberg problem is known to decrease 
exponentially~\cite{Landau,Zener,Majorana,Stuckelberg}. Suzuki and Okada~\cite{Suzuki}, however, 
calculated numerically the residual energy, the difference between the energy expectation 
$E(s)=\bra{\psi(s)} H(s)\ket{\psi(s)}$ and the instantaneous ground energy $e_{-}(s)$, for 
a modified Landau-Zener-Majorana-St\"ukelberg problem. They showed the transition of 
the residual energy from exponential decay only for short running time to the inverse-square decay
for long running time. The similar result was obtained by Rezakhani {\it et al.}~\cite{Rezakhani}.
Note $1/T^2$ decay was reported for the simulated annealing system by Santoro 
{\it et al.}~\cite{Santoro} and for adiabatic quantum teleportation by Oh {\it et al.}~\cite{Oh13}. 
As illustrated in Fig.~\ref{Fig3}, we calculate numerically the 
non-adiabatic transition probability $P(1)$ as a function of running time $T$ and find
\begin{equation}
P(1) \sim
\begin{cases}
\exp\left( -A\,T\right) & \text{for\, $T < T_c$}\\[10pt]
B/{T^2}                & \text{for\, $T > T_c$}
\end{cases}\,.
\label{transition_prob}
\end{equation}
The coefficients $A$, $B$, and the transition time $T_c$ depend on the system size $N$,
as shown in Fig.~\ref{Fig4}. The numerical data show $A\sim \pi/4N$ and $B\sim 4/N$.
The critical running time $T_c$ can be defined by a solution of the transcendental equation 
$e^{-A\,T} = B/T^2$ in Eq.~(\ref{transition_prob}). It is given by 
\begin{equation}
T_c = -\frac{2}{A} W_{-1}\left(-\frac{A\sqrt{B}}{2}\right) 
  \sim \frac{8N}{\pi} W_{-1}\left(-\frac{\pi}{4\sqrt{N^3}}\right)\,,
\label{Tc_Lambert}
\end{equation}
where $W_{-1}$ is the lower branch of the Lambert $W$ function~\cite{Lambert,Corless}.

\paragraph{Transitionless driving--} 
When the time-dependent Schr\"odinger equation is transformed to the adiabatic frame,
it is clearly seen why the non-adiabatic transition happens. Demirplak and Rice~\cite{Demirplak}, 
and Berry~\cite{Berry} showed that a time-dependent Hamiltonian $H_D(t)$, called 
the counter or transitionless driving term, in addition to the original time-dependent Hamiltonian 
makes a quantum state follow the original adiabatic state exactly. The main idea is to make 
a driving Hamiltonian cancel the non-adiabatic term seen in the adiabatic frame. The driving 
Hamiltonian $H_{D}(t)$ for Hamiltonian~(\ref{Hamil_D}) reads
\begin{align}
H_D(t) = i\hbar\,\frac{\partial U^{\dag}(t)}{\partial t}\,U(t)
=  -\hbar\,\frac{\dot{\theta}}{2}\,\sigma_y 
\label{Hamil_Driving}
\end{align}
where the unitary operator $U(t)$ is composed of instantaneous eigenstates $\ket{e_{\pm}(t)}$
\begin{align}
U(t) = \left[ 
       \begin{array}{rr}
       \cos\frac{\theta}{2} & -\sin\frac{\theta}{2}\\[10pt]
       \sin\frac{\theta}{2} &  \cos\frac{\theta}{2}
       \end{array}
       \right]\,,
\end{align}
and $\dot{\theta} = \frac{d\theta}{ds}\frac{ds}{dt} = \frac{1}{T}\frac{d\theta}{ds}$. 
Note Pauli operator $\sigma_y$ is represented by
$\sigma_y= -i\ketbra{w}{w_{\perp}} + i \ketbra{w_{\perp}}{w}$.
For linear interpolation, one has 
$\dot{\theta}(t) = 2\frac{\sqrt{N-1}}{NT}\left[(1-2s)^2 + \frac{4}{N}(1-s)s\right]^{-1}$.
As expected, the driving Hamiltonian goes to zero in the adiabatic limit, $T\gg 1$.

\begin{figure}[htbp]
\includegraphics[scale = 1.0,angle=0]{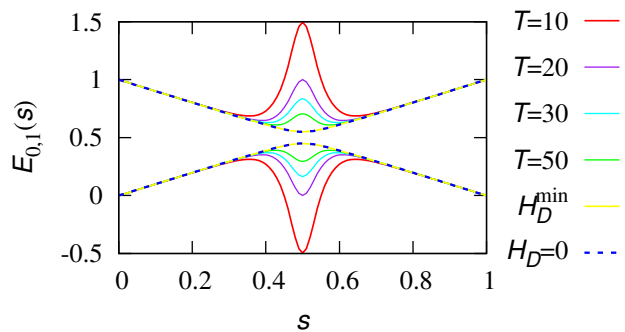}
\caption{(color online). Instantaneous eigenvalues $E_{0,1}(s)$ of $H(s) + H_D(s)$ as a function of $s$ 
for $T=10,20,30,50$, $H_D^{\rm min}$ at $T=10$, and $H_D=0$. The size of the system is $N=100$.}
\label{Fig5}
\end{figure}

\begin{figure}[htbp]
\includegraphics[scale=1.0, angle=0]{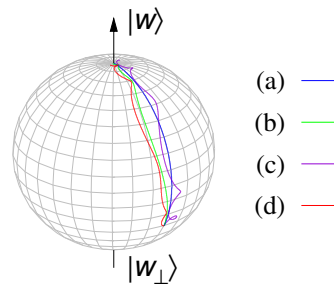}
\caption{(color online). Trajectories of Bloch vectors on a Bloch sphere
when the quantum evolution is driven (a) by adiabatically or exactly $H_D(t)$, 
(b) by $H_D^{\min}$, (c) by $H_D^{\max}$, and (d) without driving.
Here $N=4$ and $T=10$ are taken.}
\label{Fig6}
\end{figure}

\begin{figure}[htbp]
\includegraphics[scale=1.0, angle=0]{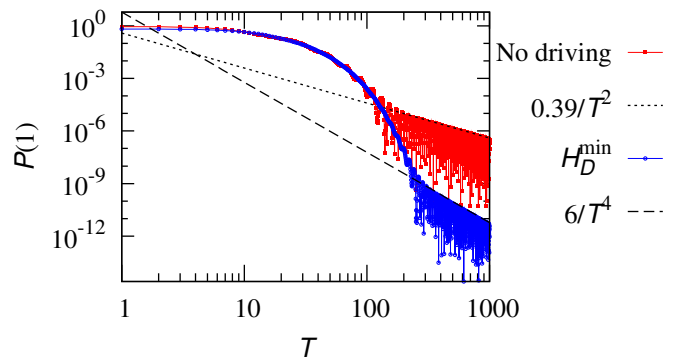}
\caption{(color online). Non-adiabatic transition probability $P(1)$ as a function of running time 
$T$ with $H_D^{\min}$ (blue) and without driving Hamiltonian (red). Here $N=10$ is taken.}
\label{Fig7}
\end{figure}

While the driving Hamiltonian $H_D(t)$ makes a quantum state evolve exactly along the longitudinal 
line (adiabatic path) regardless of $T$, it seems to be difficult to control the strength 
$\dot{\theta}$ even in linear interpolation case. So, we investigate whether an approximate 
but constant driving Hamiltonian, instead of the exact time-dependent driving 
Hamiltonian~(\ref{Hamil_Driving}), could reduce some errors. 
We consider two constant driving Hamiltonians which are the minimum and 
maximum values of $H_D$, respectively
\begin{align}
H_D^{\min} = -\frac{\hbar\sqrt{N-1}}{NT}\,\sigma_y\,,\;\;
H_D^{\max} = -\frac{\hbar\sqrt{N-1}}{T}\,\sigma_y\,.
\end{align}
Fig.~\ref{Fig5} shows how the instantaneous eigenvalues change when the driving Hamiltonian $H_D(s)$ is 
added to $H(s)$. The role of $H_D$ is to make the gap at the avoided crossing wider. While 
the approximate driving Hamiltonian $H_D^{\rm min}$ seems to make a very little change in adiabatic energy
levels and the trajectory as shown in Fig.~\ref{Fig6}, it produces drastic change in 
the non-adiabatic transition probability for long running time, from $O(1/T^2)$ to $O(1/T^4)$ 
as depicted in Fig.~\ref{Fig7}. Let take a close look at it in connection with the adiabatic condition
\begin{align}
T\gg \frac{\max_{s} |\bra{e_{+}(s)}\frac{d H}{ds}\ket{e_{-}(s)}|}{\min_s\Delta E(s)^2}\,,
\label{Adiabatic_condition}
\end{align}
where $\Delta E$ is the energy gap.
For two Hamiltonians $H(s)$ and $H(s) + H_{\min}^D$ with $T=10$, while the numerators in 
Eq.~(\ref{Adiabatic_condition}) are same, the denominators change slightly, to be more specific, 
from $0.01$ to $0.010396$. Although the right-hand side of the inequality~(\ref{Adiabatic_condition})
changes very little, $P(1)$ for long running time changes from the inverse square to 
fourth power decays. Note that $H_D^{\rm min}$ also reduces $P(1)$ for short running time. 

\paragraph{Conclusion--} 
We studied quantum dynamics of Grover's adiabatic search algorithm as a time-dependent two-level 
system. The transition from the non-adiabatic and adiabatic quantum evolutions were visualized 
by changes in trajectories of Bloch vectors on a Bloch sphere. We found a drastic change in 
the non-adiabatic transition probability from well-known exponential decay for short running 
time to the inverse-square decay for longer running time. The dependence of the critical 
running time on the problem size is obtained with Lambert $W$ function. We showed an approximate 
but constant driving Hamiltonian could reduce the non-adiabatic transition probability significantly 
which becomes the inverse fourth power decay for long running time. It would be interesting to see 
whether the results obtained in this paper could be applied to other quantum system, for example, 
a quantum Ising model~\cite{Campo}, or quantum optimization problems~\cite{Boixo}. 
While our results was obtained by numerical calculations, 
it would be interesting to seek an exact analytic solution.

\vfill

\begin{thebibliography}{99}
%
\bibitem{Grover97} Lov K. Grover, \prl, {\bf 79}, 325 (1997). 
\bibitem{Farhi} E. Farhi, J. Goldstone, S. Gutmann, J. Lapan, A. Lundgren, and D. Preda,
       Science {\bf 292}, 472 (2001).
\bibitem{Vandam} W. van Dam, M. Mosca, and U. Vazirani, Proceedings of the 42nd Annual Symposium on 
        Foundations of Computer Science, p. 279-287 (2001).
\bibitem{Aharonov} D. Aharonov, W. Van Dam, J. KEPME, Z. Landau, S. Lloyd, and O. Regev,
        SIAM J. Comput. {\bf 37}, 166 (2007).
\bibitem{Roland03} J. Roland and N.~J. Cerf, \pra\ {\bf 68}, 062311 (2003); {\it ibid}, 062312 (2003).
\bibitem{Messiah} A. Messiah, {\it Quantum Mechanics}
        (North-Holland, Amsterdam, 1963).
\bibitem{Schaller} G. Schaller, S. Mostame, R. Sch\"utzhold, \pra\ {\bf 73}, 062307 (2006).
%
\bibitem{Horn} R. A. Horn and C. R. Johnson, {\it Matrix Analysis} 
       (Cambridge Univ. Press, Cambridge, 1990), p. 39.
%
\bibitem{Betz} V. Betz and S. Teufel, in Lect.  Notes Phys. {\bf 690}, 19 (2006).
%
\bibitem{Landau} L.~D. Landau, Physics of the Soviet Union {\bf 2}, 46 (1932).
\bibitem{Zener} C.~M. Zener, Proc. R. Soc. London Ser. A {\bf 137}, 696 (1932).
\bibitem{Majorana} E. Majorana, Nuovo Cimento {\bf 9}, 43 (1932).
\bibitem{Stuckelberg} E. C. G. St\"uckelberg, Helv. Phys. Acta {\bf 5}, 369 (1932).
%
\bibitem{Suzuki} S. Suzuki and M. Okada, in Lect. Notes Phys. {\bf 679}, 207 (2005).
\bibitem{Rezakhani} A. T. Rezakhani, A. K. Pimachev, and D. A. Lidar \pra\ {\bf 82}, 052305 (2010).
\bibitem{Santoro} G.~E. Santoro, R. Marto\v{n}\'{a}k, E. Tosatti, and R. Car, Science {\bf 295}, 2427 (2002).
\bibitem{Oh13} S. Oh, Y.-P. Shim, J. Fei, M. Friesen, and X. Hu, \pra\ {\bf 87}, 022332 (2013).
%
\bibitem{Lambert} J. H. Lambert, Acta Helvetica, Physico-mathematico-anatomico-13botanico-medica {
       \bf 3}, 128 (1758).
\bibitem{Corless} R.~M. Corless, G.~H. Gonnet, D.~E.~G. Hare, D.~J. Jeffrey, and D.~E. Knuth, 
       Adv. in Comp. Math., {\bf 5} 329 (1996).
%
\bibitem{Demirplak} M. Demirplak and S. A. Rice, J. Phys. Chem. A {\bf 107}, 9937 (2003).
\bibitem{Berry} M. V. Berry, J. Phys. A: Math. Theor. {\bf 42}, 365303 (2009).
%
\bibitem{Campo} A. del Campo, M. M. Rams, and W. H. Zurek, \prl\ {\bf 109}, 115703 (2012).
\bibitem{Boixo} S. Boixo, T. Albash, F. M. Spedalieri, N. Chancellor, and D. A. Lidar,
       Nat. Commun. {\bf 4}, 3067 (2013).
\end{thebibliography}
\end{document}